\documentclass[11pt]{article}
\usepackage{amssymb, amsthm, amsmath}
\usepackage{bm,array}
\usepackage{graphicx}
\usepackage{setspace}
\usepackage{enumerate}
\usepackage{subfigure}
\usepackage[font=small,labelfont=bf]{caption}
\usepackage{natbib}
\usepackage{graphicx}
\usepackage{epstopdf}
\usepackage{epsfig}
\newcommand{\beq}{ \begin{equation}}
\newcommand{\eeq}{ \end{equation}}
\newcommand{\beqn}{ \begin{eqnarray}}
\newcommand{\eeqn}{ \end{eqnarray}}
\newcommand{\mat}[1]{\mbox{\boldmath{$#1$}}}

\begin{document}
\newcolumntype{C}{>{\centering\arraybackslash}p{2em}}
%%%%%%%%%%%%%%%%%%%%%%%%%%%%%%%%%%%%%%%%%%
%             Title page                 %
%%%%%%%%%%%%%%%%%%%%%%%%%%%%%%%%%%%%%%%%%%
% Make title, author, date
\title{\textbf{A nonparametric Bayesian test of dependence}}
%% "footnote" and "~"
%% "footnotemark" because two aithors are sharing the same footnote
\author{\textbf{Yimin Kao, ~ Brian J. Reich, ~ Howard D. Bondell}\\
\emph{\footnotesize{Department of Statistics, North Carolina State
University, Raleigh, NC}}}
%% if no "date", the default is today's date
%\date{September, 24, 2012}
\maketitle

% No number subscript on the first page
\thispagestyle{empty}

%%%%%%%%%%%%%%%%%%%%%%%%%%%%%%%%%%%%%%%%%%
%             abstract                   %
%%%%%%%%%%%%%%%%%%%%%%%%%%%%%%%%%%%%%%%%%%
\begin{abstract}
%% "noindent", "sc"
In this article, we propose a new method for the fundamental task of
testing for dependence between two groups of variables. The response
densities under the null hypothesis of independence and the
alternative hypothesis of dependence are specified by nonparametric
Bayesian models. Under the null hypothesis, the joint distribution
is modeled by the product of two independent Dirichlet Process
Mixture (DPM) priors; under the alternative, the full joint density
is modeled by a multivariate DPM prior. The test is then based on
the posterior probability of favoring the alternative hypothesis.
The proposed test not only has good performance for testing linear
dependence among other popular nonparametric tests, but is also
preferred to other methods in testing many of the nonlinear
dependencies we explored. In the analysis of gene expression data,
we compare different methods for testing pairwise dependence between
genes. The results show that the proposed test identifies some
dependence structures that are not detected by other tests.

\end{abstract}

%%%%%%%%%%%%%%%%%%%%%%%%%%%%%%%%%%%%%%%%%%%%%%%%%%%%%%%%%%%%%%%%%%%%%%%%%%%%%%%%
%%%%%%%%%%%%%%%%%%%%%%%%%%%%%%%%%%%%%%%%%%%%%%%%%%%%%%%%%%%%%%%%%%%%%%%%%%%%%%%%
%%%%%%%%%%%%%%%%%%%%%%%%%%%%%%%%%%%%%%%%%%%%%%%%%%%%%%%%%%%%%%%%%%%%%%%%%%%%%%%%
%%%%%%%%%%%%%%%%%%%%%%%%%%%%%%%%%%%%%%%%%%%%%%%%%%%%%%%%%%%%%%%%%%%%%%%%%%%%%%%%
%%%%%%%%%%%%%%%%%%%%%%%%%%%%%%%%%%%%%%%%%%%%%%%%%%%%%%%%%%%%%%%%%%%%%%%%%%%%%%%%
%%%%%%%%%%%%%%%%%%%%%%%%%%%%%%%%%%%%%%%%%%%%%%%%%%%%%%%%%%%%%%%%%%%%%%%%%%%%%%%%

% Start a new page
\newpage

%%%%%%%%%%%%%%%%%%%%%%%%%%%%%%%%%%%%%%%%%%%%%%%%%%
%                Main document                   %
%%%%%%%%%%%%%%%%%%%%%%%%%%%%%%%%%%%%%%%%%%%%%%%%%%
% Tells Latex that this is the first page to start count
\setcounter{page}{1}

% "label" is really convenient
\section{Introduction}\label{s:intro}
~~~~A fundamental task in statistics is to determine whether two
groups of variables are dependent. For example, in genomic analysis,
we might want to test whether two groups of genes are associated to
identify dependence between genetic pathways. In the brain imaging
research, we may want to discover whether sets of voxels from
different parts of the brain are related to explore functional
connectivity. In general, high-dimensional data analysis can be
simplified by identifying sets of independent variables.\\
%%%%%%%%%%%%%%%%%%%%%%%%%%%%%%%%%%%%%%%%%%%%%%%%%%%%%%%%%%%%%%%%%%%%%%%%%%%%%%%%
\indent Testing of dependence is often reduced to testing for linear
dependence. Pearson correlation coefficient is a classical and
widely-used method for quantifying the strength of linear dependence
between two univariate variables. Spearman's rank correlation
coefficient \citep{Spearman} is a ranked-based version of Pearson
correlation coefficient which quantifies monotone correlation. Tests
based on correlation are powerful for testing
specific types of association, but lose power for other general types.\\
%%%%%%%%%%%%%%%%%%%%%%%%%%%%%%%%%%%%%%%%%%%%%%%%%%%%%%%%%%%%%%%%%%%%%%%%%%%%%%%%
\indent For testing more general associations, the $\chi^2$ test of
independence and Hoeffding's test of independence
\citep{Dstatistics} are two classical nonparametric methods. These
tests are based on partitioning data into a contingency table. The
main drawback for $\chi^2$ test is that the result is sensitive to
the way the data are partitioned. Several approximations of the test
statistics of the Hoeffding's test are studied: \cite{Hoeffding2}
introduce an approximation by the concordances and discordances of a
$2\times 2$ contingency tables, and \cite{Wilding2008160} propose an
approximation by using two Weibull extensions. A relation between
the Hoeffding's test and the $\chi^2$ test statistics was noted by
\cite{modHoeff}, and they also suggested extending the idea of
\cite{Hoeffding2} to a $k\times k$ contingency tables, for $k>2$.
More recent methods related to the Hoeffding's test have been
proposed by \cite{HHG} and \cite{DDP}. Both of these tests are
consistent under general types of associations. Other methods for
testing for independence include the distance correlation test of
\cite{dCov} and the maximal information coefficient of \cite{MIC}.
Both the tests of \cite{HHG} and \cite{dCov} can be extended to
higher dimensions for testing joint independence of two or more
random vectors. Several Bayesian methods are available for testing
of independence. The simplest test of linear dependence between two
univariate random variables can be achieved by fitting a linear
model and inspecting the posterior distribution of the correlation
coefficient. Other methods were proposed for testing of independence
based on a contingency table
\citep{BayesContingency,BayesContingency3,BayesContingency2}.\\
%%%%%%%%%%%%%%%%%%%%%%%%%%%%%%%%%%%%%%%%%%%%%%%%%%%%%%%%%%%%%%%%%%%%%%%%%%%%%%%%
\indent In this article, we propose a nonparametric Bayesian test of
independence between two groups of variables. We test the null
hypothesis of independence and the alternative hypothesis of
dependence. We specify nonparametric Bayesian models for the
response density under both hypotheses. Under the null hypothesis,
the joint distribution is taken to be the product of two independent
densities, both with nonparametric priors; under the alternative,
the full joint density has a nonparametric prior. The test is based
on the posterior probability of the alternative hypothesis. By
specifying nonparametric Bayesian models under each hypothesis, we
obtain an extremely flexible test which can capture both linear and
complex nonlinear relationships between groups of variables.\\
%%%%%%%%%%%%%%%%%%%%%%%%%%%%%%%%%%%%%%%%%%%%%%%%%%%%%%%%%%%%%%%%%%%%%%%%%%%%%%%%
\indent The remainder of the article proceeds as follows. In Section
\ref{s:sm}, we introduce the statistical algorithm. The details of
the reversible jump MCMC algorithm use to compute the posterior
probability of the alternative hypothesis are provided in Section
\ref{s:Cd}. In Section \ref{s:sim}, we present a simulation study to
compare the power of the proposed test with other tests of linear
and nonlinear relationships. The method is illustrated using a
genetic data analysis in Section \ref{s:real}. Section
\ref{s:conclusion} concludes.

%%%%%%%%%%%%%%%%%%%%%%%%%%%%%%%%%%%%%%%%%%%%%%%%%%%%%%%%%%%%%%%%%%%%%%%%%%%%%%%%
%%%%%%%%%%%%%%%%%%%%%%%%%%%%%%%%%%%%%%%%%%%%%%%%%%%%%%%%%%%%%%%%%%%%%%%%%%%%%%%%
%%%%%%%%%%%%%%%%%%%%%%%%%%%%%%%%%%%%%%%%%%%%%%%%%%%%%%%%%%%%%%%%%%%%%%%%%%%%%%%%
%%%%%%%%%%%%%%%%%%%%%%%%%%%%%%%%%%%%%%%%%%%%%%%%%%%%%%%%%%%%%%%%%%%%%%%%%%%%%%%%
%%%%%%%%%%%%%%%%%%%%%%%%%%%%%%%%%%%%%%%%%%%%%%%%%%%%%%%%%%%%%%%%%%%%%%%%%%%%%%%%
%%%%%%%%%%%%%%%%%%%%%%%%%%%%%%%%%%%%%%%%%%%%%%%%%%%%%%%%%%%%%%%%%%%%%%%%%%%%%%%%

\section{Statistical model}\label{s:sm}
~~~~Let $\mat{X}_1\in\mathbb{R}^{D_1}$ and
$\mat{X}_2\in\mathbb{R}^{D_2}$ be random vectors in $D_1$ and $D_2$
dimensions, respectively, and denote
$\mat{X}=(\mat{X}_1,\mat{X}_2)$. The objective is to test whether
$\mat{X}_1$ and $\mat{X}_2$ are independent. The hypotheses are
\beqn
\mbox{H}_0&:&\mbox{$\mat{X}_1$ and $\mat{X}_2$ are independent and $f(\mat{X})=f_1(\mat{X}_1)f_2(\mat{X}_2)$}\nonumber\\
\mbox{H}_1&:&\mbox{$\mat{X}_1$ and $\mat{X}_2$ are dependent and
$f(\mat{X})$ cannot be factorized}\nonumber\eeqn In other words,
when they are independent, the joint density can be factorized as
the product of
two lower-dimensional densities.\\
%%%%%%%%%%%%%%%%%%%%%%%%%%%%%%%%%%%%%%%%%%%%%%%%%%%%%%%%%%%%%%%%%%%%%%%%%%%%%%%%
\indent Under both hypotheses, the densities are modeled using
Dirichlet process mixture (DPM) prior. Under $\mbox{H}_0$,
$f_1(\mat{X}_1)$ and $f_2(\mat{X}_2)$ follow independent DPM priors;
under $\mbox{H}_1$ when $\mat{X}_1$ and $\mat{X}_2$ are not
independent, the joint distribution is assumed to follow a DPM
prior. The following subsections describe the independent and joint
DPM priors.

%%%%%%%%%%%%%%%%%%%%%%%%%%%%%%%%%%%%%%%%%%%%%%%%%%%%%%%%%%%%%%%%%%%%%%%%%%%%%%%%
%%%%%%%%%%%%%%%%%%%%%%%%%%%%%%%%%%%%%%%%%%%%%%%%%%%%%%%%%%%%%%%%%%%%%%%%%%%%%%%%

\subsection{The independent DPM prior}
~~~~When $\mat{X}_1$ and $\mat{X}_2$ are independent,
$f_j(\mat{X}_j)$, $j=1,2$, are assumed to follow the DPM prior
independently. The DPM prior can be written as the infinite mixture
\beq\label{indepDPM}
f_j(\mat{X}_j)=\sum_{l=1}^{\infty}w_{lj}\phi_j(\mat{X}_j~|~\mat{\mu}_{lj},\mat{\Sigma}_j),\eeq
where $w_{lj}$ is the mixture weight, $\phi_j$ is assigned to be the
$D_j-$dimensional multivariate normal distribution (MVN) in this
analysis, $\mat{\mu}_{lj}$ is the mean vector of the $l^{th}$
mixture
component, and $\mat{\Sigma}_j$ is the covariance matrix.\\
%%%%%%%%%%%%%%%%%%%%%%%%%%%%%%%%%%%%%%%%%%%%%%%%%%%%%%%%%%%%%%%%%%%%%%%%%%%%%%%%
\indent The mixture weights $w_{lj}$ are modeled by the
stick-breaking construction with concentration parameter $d_j$. The
weights $w_{lj}$ are modeled in terms of latent
$v_{lj}\sim\mbox{Beta}(1,d_j)$. The first weight is $w_{1j}=v_{1j}$.
The remaining elements are modeled as
$w_{lj}=v_{lj}\prod_{i=1}^{l-1}(1-v_{ij})$, where
$\prod_{i=1}^{l-1}(1-v_{ij})=1-\sum_{i=1}^{l-1}w_{ij}$ is the
remaining probability after accounting first $l-1$ mixture weights.
The number of mixture components is truncated by a sufficiently
large number $K$ (i.e. $l=1,...,K$), where the last term $v_K$ is
fixed to be 1 to ensure that $\sum_{l=1}^Kw_{lj}=1$.\\
%%%%%%%%%%%%%%%%%%%%%%%%%%%%%%%%%%%%%%%%%%%%%%%%%%%%%%%%%%%%%%%%%%%%%%%%%%%%%%%%
\indent The mean vectors $\mat{\mu}_{lj}$ have priors
$\mat{\mu}_{lj}\sim\mbox{MVN}(\mat{0},\mat{\Omega}_j)$. The
covariance matrices  $\mat{\Sigma}_j$ and $\mat{\Omega}_j$ are
parameterized as $\mat{\Sigma}_j=r\mat{S}_j$, and
$\mat{\Omega}_j=(1-r)\mat{S}_j$. Under this model, $\mat{S}_j$ is
the covariance matrix for $\mat{X}_j$ marginally over the mixture
means $\mat{\mu}_{lj}$, and $r$ is the proportion of the total
variance attributed to the variance within each mixture component.
The marginal covariance $\mat{S}_j$ is assigned to have inverse
Wishart prior distribution, and to facilitate computing, the prior
of $r$ is a discrete uniform distribution with support
$r\in\{0,0.01,...,1\}$. The concentration parameter $d_j$ has prior
distribution $\mbox{Gamma}(a,b)$.

%%%%%%%%%%%%%%%%%%%%%%%%%%%%%%%%%%%%%%%%%%%%%%%%%%%%%%%%%%%%%%%%%%%%%%%%%%%%%%%%
%%%%%%%%%%%%%%%%%%%%%%%%%%%%%%%%%%%%%%%%%%%%%%%%%%%%%%%%%%%%%%%%%%%%%%%%%%%%%%%%

\subsection{The joint DPM prior}
~~~~When $\mat{X}_1$ and $\mat{X}_2$ are not independent,
$f(\mat{X})$ is assumed to follow the joint DPM
prior\beq\label{jointDPM}
f(\mat{X})=\sum_{l=1}^{\infty}w_{l}\phi(\mat{X}~|~\mat{\mu}_{l},\mat{\Sigma}),\eeq
where $w_{l}$ is the mixture weight, $\phi$ is the
$(D_1+D_2)-$dimensional MVN distribution, $\mat{\mu}_{l}$ is the
mean vector of the $l^{th}$ mixture component, and $\mat{\Sigma}$ is
the covariance matrix. The number of mixtures is truncated by the
same number $K$ as in the independent model. The mixture weights
$w_{l}$ are again modeled by the stick-breaking algorithm with
concentration parameter $d$. The mean vectors $\mat{\mu}_l$ have
priors $\mat{\mu}_l\sim \phi(\mat{0},\mat{\Omega})$. The covariance
matrices $\mat{\Sigma}$ and $\mat{\Omega}$ are modeled as
$\mat{\Sigma}=r\mbox{diag}(\mat{S})$ and
$\mat{\Omega}=(1-r)\mat{S}$, where $\mat{S}$ is the covariance
matrix for $\mat{X}$, and $\mbox{diag}(\mat{S})$ is the diagonal
form of $\mat{S}$. In other words, under the joint DPM prior, we
assign non-diagonal structure for the $\mat{\Omega}$, and diagonal
structure for the $\mat{\Sigma}$. We found that diagonalizing
$\mat{\Sigma}$ greatly improved computational stability . The priors
for $\mat{S}$, $r$, and $d$ are the same as in the independent DPM
prior.

%%%%%%%%%%%%%%%%%%%%%%%%%%%%%%%%%%%%%%%%%%%%%%%%%%%%%%%%%%%%%%%%%%%%%%%%%%%%%%%%
%%%%%%%%%%%%%%%%%%%%%%%%%%%%%%%%%%%%%%%%%%%%%%%%%%%%%%%%%%%%%%%%%%%%%%%%%%%%%%%%

\subsection{Bayesian test of independence}\label{ss:transition}~~~~The Bayesian hypothesis test of
independence is based on the Bayes factor (BF)\beq
\mbox{BF}=\frac{P(\mbox{H}_1~|~\mat{X})/P(\mbox{H}_0~|~\mat{X})}{P(\mbox{H}_1)/P(\mbox{H}_0)}=\frac{P(\mat{X}~|~\mbox{H}_1)}{P(\mat{X}~|~\mbox{H}_0)}.\eeq
The null is rejected if $\mbox{BF}>T$, where $T$ is a threshold
parameter. The threshold parameter $T$ can be chosen based on rules
of thumb about the weight of evidence favoring $\mbox{H}_1$. For
example, \cite{BF} suggest that $\mbox{BF}=10$ is a strong evidence
for $\mbox{H}_1$. Alternatively, in the simulation study in Section
\ref{s:sim}, we select $T$ to control the Type I error rate. In the
analysis of genetic data in Section \ref{s:real}, multiple tests are
performing simultaneously, therefore we select $T$ to control the
Bayesian false discovery rate.

%%%%%%%%%%%%%%%%%%%%%%%%%%%%%%%%%%%%%%%%%%%%%%%%%%%%%%%%%%%%%%%%%%%%%%%%%%%%%%%%
%%%%%%%%%%%%%%%%%%%%%%%%%%%%%%%%%%%%%%%%%%%%%%%%%%%%%%%%%%%%%%%%%%%%%%%%%%%%%%%%
%%%%%%%%%%%%%%%%%%%%%%%%%%%%%%%%%%%%%%%%%%%%%%%%%%%%%%%%%%%%%%%%%%%%%%%%%%%%%%%%
%%%%%%%%%%%%%%%%%%%%%%%%%%%%%%%%%%%%%%%%%%%%%%%%%%%%%%%%%%%%%%%%%%%%%%%%%%%%%%%%
%%%%%%%%%%%%%%%%%%%%%%%%%%%%%%%%%%%%%%%%%%%%%%%%%%%%%%%%%%%%%%%%%%%%%%%%%%%%%%%%
%%%%%%%%%%%%%%%%%%%%%%%%%%%%%%%%%%%%%%%%%%%%%%%%%%%%%%%%%%%%%%%%%%%%%%%%%%%%%%%%

\section{Computing details}\label{s:Cd}
~~~~Computing the Bayes factor requires computing the posterior
probability of each hypothesis. This is accomplished using a
reversible jump MCMC (RJMCMC) algorithm as described below.
%%%%%%%%%%%%%%%%%%%%%%%%%%%%%%%%%%%%%%%%%%%%%%%%%%%%%%%%%%%%%%%%%%%%%%%%%%%%%%%%
%%%%%%%%%%%%%%%%%%%%%%%%%%%%%%%%%%%%%%%%%%%%%%%%%%%%%%%%%%%%%%%%%%%%%%%%%%%%%%%%

\subsection{Reparameterization and hyperparameters}
~~~~The updating algorithm of the DPM prior is facilitated by
introducing the equivalent clustering model. The mixture form in
(\ref{indepDPM}) can be written as
$$f_j(\mat{X}_j~|~g_j=l)=\phi_j(\mat{X}_j~|~\mat{\mu}_{lj},\mat{\Sigma}_j),$$ which draws an
auxiliary cluster label $g_j\in\{1,...,K\}$ with $P(g_j=l)=w_{lj}$.
Similarly, the model in (\ref{jointDPM}) is equivalent to
$$f(\mat{X}~|~g=l)=\phi(\mat{X}~|~\mat{\mu}_{l},\mat{\Sigma}),$$ with cluster label $g$ and
$P(g=l)=w_{l}$. Under the clustering model, the full conditionals of
all the parameters
are conjugate.\\
%%%%%%%%%%%%%%%%%%%%%%%%%%%%%%%%%%%%%%%%%%%%%%%%%%%%%%%%%%%%%%%%%%%%%%%%%%%%%%%%
\indent In addition, we introduce model indicator parameter $M$,
where
\[M\in\left\{\begin{array}{ll}
         I & \mbox{if $\mat{X}_1$ and $\mat{X}_2$ are independent ($\mbox{H}_0$ is true)}\\
         J & \mbox{if $\mat{X}_1$ and $\mat{X}_2$ are not independent ($\mbox{H}_1$ is true)}.\end{array} \right.\]
Under each MCMC step, we propose a new indicator $M^{'}$ in the
Markov chain, and decide whether to accept the new status $M^{'}$.
The probability $P(\mbox{H}_1~|~\mat{X})$ is then approximated by
$\sum_{i=1}^NI(M^{(i)}=J)/N$, where $N$ is the number of MCMC
samples and $M^{(i)}$ is the model status for the $i^{th}$ MCMC
sample.\\
%%%%%%%%%%%%%%%%%%%%%%%%%%%%%%%%%%%%%%%%%%%%%%%%%%%%%%%%%%%%%%%%%%%%%%%%%%%%%%%%
\indent Throughout this article, we let the number of mixture
components truncated at $K=20$ and the hyperparameters in the
stick-breaking procedure $(a,b)$ are fixed under different sample
sizes $n$ as presented in Table \ref{abtable}.
\begin{table}[h]
\caption{Hyperparameters $(a,b)$ under different sample sizes
$n$.}\label{abtable}
\begin{center}
\begin{tabular}
{|c|c|c|} \hline $n$ & $a$ & $b$\\
\hline 100                 & 1.5 & 2.5\\
       200                 & 1.0 & 4.0\\
       300                 & 1.0 & 4.5\\
       500                 & 0.8 & 4.6\\ \hline
\end{tabular}
\end{center}
\end{table}
%%%%%%%%%%%%%%%%%%%%%%%%%%%%%%%%%%%%%%%%%%%%%%%%%%%%%%%%%%%%%%%%%%%%%%%%%%%%%%%%
%%%%%%%%%%%%%%%%%%%%%%%%%%%%%%%%%%%%%%%%%%%%%%%%%%%%%%%%%%%%%%%%%%%%%%%%%%%%%%%%

\subsection{Pseudo code for the DPM test of independence algorithm}\label{ss:pseudo}
~~~~Let $\mat{\Theta}_M$ denote the DPM parameters
($\mat{\Theta}_M=\{\mat{\mu}_{11},...,\mat{\mu}_{K2}$, $r$,
$\mat{S}_1$, $\mat{S}_2$, $w_{11},...,w_{K2}$, $d_1,d_2$\} if $M=I$,
and $\mat{\Theta}_M=\{\mat{\mu}_{1},...,\mat{\mu}_{K}$, $r$,
$\mat{S}$, $w_{1},...,w_{K}$, $d$\} if $M=J$). The algorithm of the
DPM test of
independence is described as follows:\\
\textbf{Step 0}: Select initial values for $M$ and
$\mat{\Theta}_M$.\\
\textbf{Step 1}: Update $\mat{\Theta}_M$ given $M$ using the Gibbs
sampling.\\
\textbf{Step 2}: Update $M$ given the parameters $\mat{\Theta}_M$.\\
~~\textbf{Step 2.1}: Generate proposed model status $M^{'}$ with
$P(M^{'}=I)=P(M^{'}=J)=0.5$.\\
~~\textbf{Step 2.2}: If $M=M^{'}$, then so back to \textbf{Step 1}.\\
~~\textbf{Step 2.3}: If $M=I$ and $M^{'}=J$, then propose
$\mat{\Theta}_{M^{'}}$ required for the joint DPM prior ($\mbox{H}_1$).\\
~~\textbf{Step 2.4}: If $M=J$ and $M^{'}=I$, then propose
$\mat{\Theta}_{M^{'}}$ required for the independent DPM prior ($\mbox{H}_0$).\\
~~\textbf{Step 2.5}: Accept $M^{'}$ with probability
min$\{1,\alpha(M,M^{'})\}$.\\
\textbf{Step 3}: Back to \textbf{Step 1}.\\
The full conditionals requires for Step 1 are all standard and are
given in Appendix \ref{MCMCindep} for $M=I$, and Appendix
\ref{MCMCjoint} for $M=J$. Details on the RJMCMC steps are provided
below in Section \ref{ss:RJMCMCsteps}.

%%%%%%%%%%%%%%%%%%%%%%%%%%%%%%%%%%%%%%%%%%%%%%%%%%%%%%%%%%%%%%%%%%%%%%%%%%%%%%%%
%%%%%%%%%%%%%%%%%%%%%%%%%%%%%%%%%%%%%%%%%%%%%%%%%%%%%%%%%%%%%%%%%%%%%%%%%%%%%%%%

\subsection{Steps of the RJMCMC algorithm}\label{ss:RJMCMCsteps}
~~~~The parameter spaces under the independent and the joint DPM
priors are different, so moving between these two parameter spaces
becomes a trans-dimensional problem. Reversible jump MCMC (RJMCMC)
was first introduced by \cite{RJMCMC1}, which can be thought of as a
generalized Metropolis-Hastings algorithm for the trans-dimensional
updates.\\
%%%%%%%%%%%%%%%%%%%%%%%%%%%%%%%%%%%%%%%%%%%%%%%%%%%%%%%%%%%%%%%%%%%%%%%%%%%%%%%%
\indent Under the current model status $M$, the propose model status
$M^{'}$ is randomly assigned to be either $I$ or $J$ with acceptance
probability min$\{1,\alpha(M,M^{'})\}$, where
\beq\label{RJMCMCacc}\alpha(M,M^{'})=\frac{l_{M^{'}}\cdot\pi_{M^{'}}\cdot
q_{M~|~M^{'}}\cdot p_{M^{'}\rightarrow M}}{l_{M}\cdot\pi_{M}\cdot
q_{M^{'}~|~M}\cdot p_{M\rightarrow
M^{'}}}\left|\mathfrak{J}\right|,\eeq where $l_M$ and $\pi_M$ are
the likelihood function and the prior distribution under model $M$,
$q_{M^{'}~|~M}$ is the candidate distribution of the parameters when
proposing for model $M^{'}$ under model $M$, $p_{M\rightarrow
M^{'}}$ is the probability of proposing $M^{'}$ conditional on the
current status $M$, and $\left|\mathfrak{J}\right|$ is the Jacobian.
As $M^{'}$ is randomly picked from $\{I,J\}$, $p_{M\rightarrow
M^{'}}$ and $p_{M^{'}\rightarrow M}$ are equal in the algorithm.
Note that when $M=M^{'}$, it becomes the usual fixed-dimensional
MCMC algorithm as $\alpha(M,M^{'})=1$; when $M\neq M^{'}$, the
candidate distribution of the parameters $q$ is then for balancing
the parameter spaces
between the independent and joint models.\\
%%%%%%%%%%%%%%%%%%%%%%%%%%%%%%%%%%%%%%%%%%%%%%%%%%%%%%%%%%%%%%%%%%%%%%%%%%%%%%%%
\indent Recall that $\mat{\Theta}_M$ and $\mat{\Theta}_{M^{'}}$
denote the DPM parameters under models $M$ and $M^{'}$,
respectively, and the truncated number $K$ under both models are
assigned to be identical. We first examine the case when $X_1$ and
$X_2$ are univariate random variables ($D_1=D_2=1$) with the current
model status $M=I$, and the proposed model is $M'=J$. Denote the
covariance matrix under the joint
model as $\mat{S}=\bigl(\begin{smallmatrix} S_{11}^2&S_{11}S_{22}\rho_J\\
S_{11}S_{22}\rho_J&S_{22}^2
\end{smallmatrix} \bigr)$. We assign the $2\times K$ mean vector $\mat{\mu}$ to be the same in both the independent and joint DPM models.
Also, we assign the variances $S_{11}^2$ and $S_{22}^2$, and $r$ to
be the same across different model statuses. Therefore, this move
only requires proposing the parameters under the joint DPM prior in
(\ref{jointDPM}): the cluster label $g'_J$, $\rho'_J$, the
concentration parameter $d'_J$, and the mixture weights
$\mat{w}'_J$. The concentration parameter $d'_J$ is proposed by
$d'_J\sim\mbox{Gamma}(\bar{d_I},1)$, where $\bar{d_I}$ is the mean
of $d_I$, and then the mixture probabilities $\mat{w}'_J$ is
proposed from the stick-breaking procedure with concentration
parameter $d'_J$. The cluster label $g'_J$ is proposed from the full
conditional distribution given in Appendix \ref{fullcond}. The
details of the mapping for each parameter is described in the end of
this section.\\
%%%%%%%%%%%%%%%%%%%%%%%%%%%%%%%%%%%%%%%%%%%%%%%%%%%%%%%%%%%%%%%%%%%%%%%%%%%%%%%%
\indent Conversely, if the current model status is $M=J$ and the
proposed model status is $M'=I$, the parameters of the independent
model described in (\ref{indepDPM}) are proposed as follow: The
concentration parameter $d'_I\sim\mbox{Gamma}(d_J,1)$, and the
mixture weights $\mat{w}'_I$ are again proposed by the
stick-breaking procedure with concentration parameter $d'_I$. The
cluster label $g'_I$ is again proposed by the full conditional
distribution given in Appendix \ref{fullcond}.\\
%%%%%%%%%%%%%%%%%%%%%%%%%%%%%%%%%%%%%%%%%%%%%%%%%%%%%%%%%%%%%%%%%%%%%%%%%%%%%%%%
\indent For dimension matching under the RJMCMC algorithm, the
bijection map is described below for the case where $M=I$ and
$M'=J$. The reverse move uses the same map. Let \beqn
\theta_M&=&\{\mat{\mu},S_{11},S_{22},r,\mat{w}_I,d_I,g_I\}\nonumber\\
u&=&\{\rho'_J,\mat{w}'_J,d'_{J},g'_J\}\nonumber\\
\theta_{M'}&=&\{\mat{\mu},S_{11},S_{22},r,\mat{w}_J,\mat{d}_J,g_J,\rho_J\}\nonumber\\
u'&=&\{\mat{w}'_I,d'_{I},g'_I\}.\nonumber\eeqn Then we assign
$\mat{\Theta}_M=\{\theta_M,u\}$,
$\mat{\Theta}_{M'}=\{\theta_{M'},u'\}$. The bijection function $h$
has the form
$$h(\mat{\Theta}_M)=h(\theta_M,u)=\mat{\Theta}_{M'}=\{\theta_{M'},u'\},$$ which is
a one-to-one bijection map with: $\mat{w}_I\rightarrow\mat{w}'_I$,
$d_I\rightarrow d'_I$, $g_I\rightarrow g'_I$, $\rho'_J\rightarrow
\rho_J$, $\mat{w}'_J\rightarrow \mat{w}_J$, $d'_J\rightarrow d_J$,
and $g'_J\rightarrow g_J$. Hence, the Jacobian
$|\mathfrak{J}|=\left|\frac{\partial(\theta'_{M'},u')}{\partial(\theta_{M},u)}\right|=1$.\\
%%%%%%%%%%%%%%%%%%%%%%%%%%%%%%%%%%%%%%%%%%%%%%%%%%%%%%%%%%%%%%%%%%%%%%%%%%%%%%%%
\indent When $D_1+D_2>2$, the transition of the covariance matrices
between the independent and joint models becomes more complicated as
the off-diagonal elements are harder to propose than in the
bivariate case. One way to alleviate this concern is to assume the
covariance matrix $\mat{S}$
under the joint model is a block-diagonal matrix $\mat{S}=\bigl(\begin{smallmatrix} \mat{S}_{1}&0\\
0&\mat{S}_2
\end{smallmatrix} \bigr)$, where $\mat{S}_i$ is a $D_i\times D_i$
covariance matrix of $\mat{X}_i$ for $i=1,2$. However, in the
simulation study and the real data analysis of this article, we will
focus on the case where $X_1$ and $X_2$ are univariate random
variables.

%%%%%%%%%%%%%%%%%%%%%%%%%%%%%%%%%%%%%%%%%%%%%%%%%%%%%%%%%%%%%%%%%%%%%%%%%%%%%%%%
%%%%%%%%%%%%%%%%%%%%%%%%%%%%%%%%%%%%%%%%%%%%%%%%%%%%%%%%%%%%%%%%%%%%%%%%%%%%%%%%
%%%%%%%%%%%%%%%%%%%%%%%%%%%%%%%%%%%%%%%%%%%%%%%%%%%%%%%%%%%%%%%%%%%%%%%%%%%%%%%%
%%%%%%%%%%%%%%%%%%%%%%%%%%%%%%%%%%%%%%%%%%%%%%%%%%%%%%%%%%%%%%%%%%%%%%%%%%%%%%%%
%%%%%%%%%%%%%%%%%%%%%%%%%%%%%%%%%%%%%%%%%%%%%%%%%%%%%%%%%%%%%%%%%%%%%%%%%%%%%%%%
%%%%%%%%%%%%%%%%%%%%%%%%%%%%%%%%%%%%%%%%%%%%%%%%%%%%%%%%%%%%%%%%%%%%%%%%%%%%%%%%

\section{Simulation Study}\label{s:sim}
~~~~The simulation study focuses on testing for dependence between
two univariate variables. The objective is to compare the power of
each method under linear and nonlinear dependence. In the following
subsections, we introduce the data generation procedure, the
competing methods, and the simulation results.

%%%%%%%%%%%%%%%%%%%%%%%%%%%%%%%%%%%%%%%%%%%%%%%%%%%%%%%%%%%%%%%%%%%%%%%%%%%%%%%%
%%%%%%%%%%%%%%%%%%%%%%%%%%%%%%%%%%%%%%%%%%%%%%%%%%%%%%%%%%%%%%%%%%%%%%%%%%%%%%%%

\subsection{Data generation}\label{ss:DG}
\begin{figure}[h]
\setlength{\abovecaptionskip}{1ex} \centering
\includegraphics[height=13cm,width=15cm]{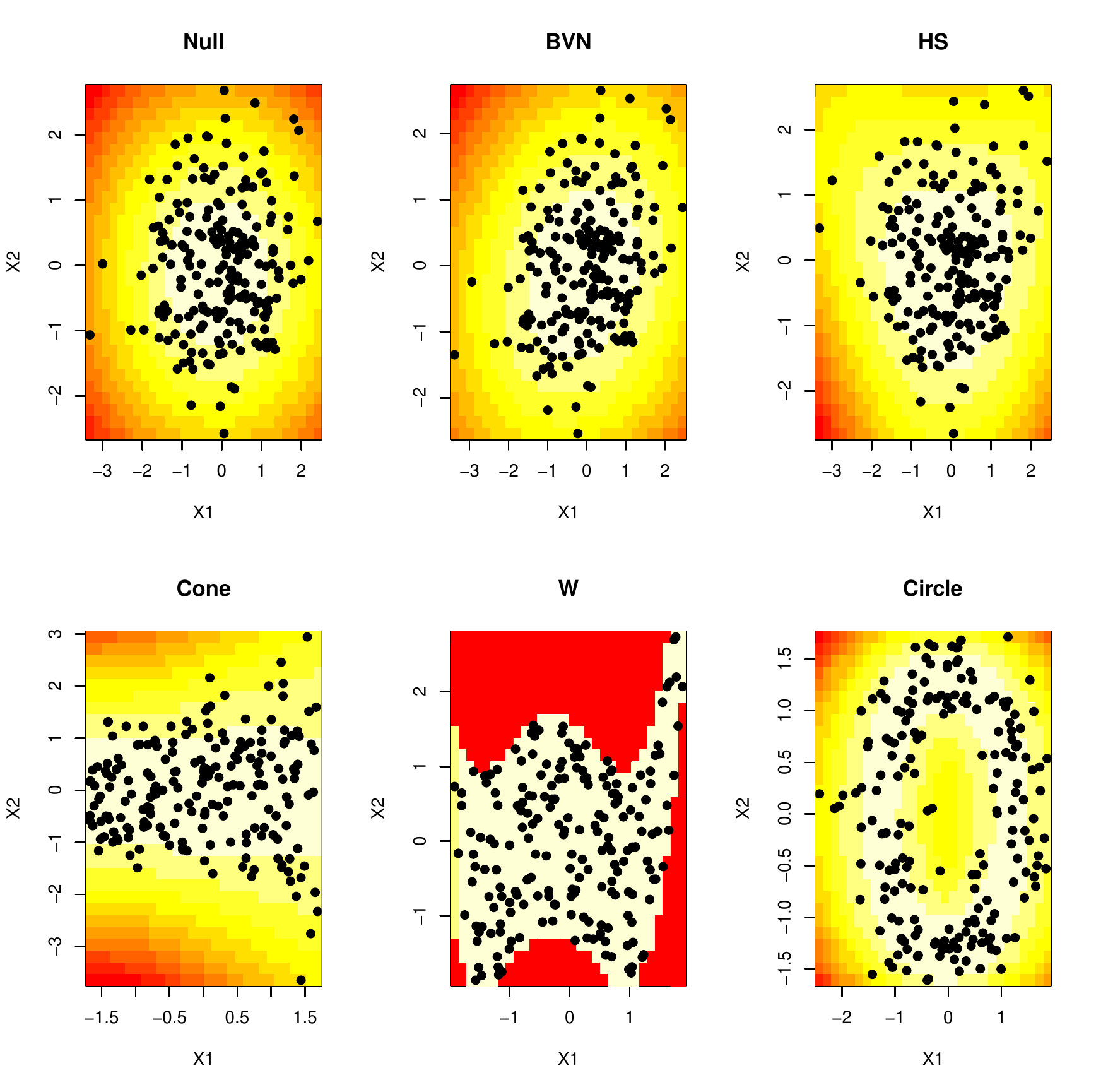}
\caption{True log density (background color) and one simulated data
set (points) for each simulation design.}\label{SimData}
\end{figure}

~~~~ The seven different types of data sets are simulated. Scenarios
5 and 6 are designed from \cite{DDP}.
\begin{enumerate}
\item Independent normal (Null): $X_j\sim\mbox{N}(0,1)$, for j=1,2.
\item Bivariate normal (BVN): $(X_1,X_2)\sim\mbox{BVN}\left[\mat{0},\bigl(\begin{smallmatrix} 1&\rho\\
\rho&1
\end{smallmatrix} \bigr)\right]$, where $\rho=0.2$.
\item Horseshoe (HS): $X_1\sim\mbox{N}(0,1)$, $X_2~|~X_1\sim\mbox{N}(\rho X_1^2,1)$, where $\rho=0.2$.
\item Cone: $X_1\sim\mbox{U}(0,1)$, $X_2~|~X_1\sim\mbox{N}\left[0,(\rho X^2_1+0.1)^2\right]$, where $\rho=0.1$.
\item W: $X_1\sim\frac{1}{n}\sum_{i=1}^{n}\mbox{U}(a_i,a_i+\frac{1}{3})$, $X_2~|~X_1\sim\mbox{U}\left[3(X_1^2-\frac{1}{2})^2,3(1+X_1^2-\frac{1}{2})\right]$,
where $a_1=-1$, $n$ is the number of samples, and
$a_i=a_{i-1}+\frac{2}{n}$, for $i>1$.
\item Circle: $(X_1,X_2)\sim\frac{1}{n}\sum_{i=1}^{n}\mbox{BVN}\left[\theta_i,\bigl(\begin{smallmatrix} \frac{1}{9}&0\\
0&\frac{1}{64}
\end{smallmatrix} \bigr)\right]$, where
$\theta_i=\left[\sin(a_i\pi),\cos(a_i\pi)\right]$, and $a_i$ is
defined as in W.
%\item 4 clouds: $(X_1,X_2)\sim\sum_{i=1}^{4}\nu_i\mbox{BVN}\left[\eta_i,\bigl(\begin{smallmatrix} \frac{1}{9}&0\\
%0&\frac{1}{9}
%\end{smallmatrix} \bigr)\right]$, where
%$(\nu_1,\nu_2,\nu_3,\mu_4)=(0.55,0.15,0.15,0.15)$, and $\left(\begin{smallmatrix} \eta_1\\
%\eta_2\\ \eta_3\\ \eta_4
%\end{smallmatrix} \right)=\left(\begin{smallmatrix} -1&1\\
%-1&-1\\ 1&1\\ 1&-1
%\end{smallmatrix} \right)$.
\end{enumerate}
~~~~Each scenario is generated with the algorithms introduced above
with sample size $n=100,200$, and 500. Then for each dimension, we
standardize the data to have mean zero and variance one. We plot the
data when $n=200$ in Figure \ref{SimData} along with the true
density. The responses are dependent for designs 2-6. Design 3-6 are
all examples of the challenging dependent but uncorrelated random
variables and thus the usual test of correlation will miss this
dependence.

%%%%%%%%%%%%%%%%%%%%%%%%%%%%%%%%%%%%%%%%%%%%%%%%%%%%%%%%%%%%%%%%%%%%%%%%%%%%%%%%
%%%%%%%%%%%%%%%%%%%%%%%%%%%%%%%%%%%%%%%%%%%%%%%%%%%%%%%%%%%%%%%%%%%%%%%%%%%%%%%%

\subsection{Methods for testing of independence}
~~~~We compare six methods in the simulation study (described in
detail in the Appendix). Each method is controlled to have type I
error rate approximately equal to 0.05.
\begin{enumerate}
%\item Bayesian linear regression (BLR): Let
%$X_1~|~\beta,\tau\sim\mbox{N}(X_2\beta,\tau)$, where $\beta$ has
%flat Normal prior distribution, and $\tau$ has flat Gamma prior
%distribution. The test of independence is whether the posterior 95\%
%credible interval of $\beta$ contains 0 or not.
\item Linear regression (LR): The model $X_2=\beta_0+\beta_1X_1+\epsilon$, $\epsilon\sim\mbox{N}(0,1)$ is fitted by least squares and the linear association is
determined by the test of $\beta_1=0$.
\item E-statistics (ES) \citep{dCov}: The testing procedure is by calculating the
distance covariance between $X_1$ and $X_2$.
\item Heller-Heller-Gorfine method (HHG) \citep{HHG}: The test statistic is
based on the sum of all likelihood ratio tests of $2\times 2$
contingency tables formed by the pairwise distances within each of
$X_1$ and $X_2$.
\item Data Derived Partitions method (DDP) \citep{DDP} with $3\times 3$ contingency tables:
The DDP method is similar to the HHG method, but only designed for
univariate random variables. The test statistic is based on the sum
of all likelihood ratio tests of $3\times 3$ contingency tables
formed by the observed values.
\item Maximal Information Coefficient method (MIC) \citep{MIC}: It is a rank-order test
statistic which is calculated from the largest achievable mutual
information under different grid sizes.
\item The DPM test of independence (DPM): The proposed test is described in
Section \ref{s:sm}. $X$ is first marginally transformed to be
standard normal distribution. The normal score transformation makes
the proposed method a distribution-free testing procedure.
Therefore, the threshold for the BF in Section \ref{s:sm} that
controls Type I error can be determined by the permutations of the
transformed data. The threshold $T$ for the Bayes factor is computed
from 300 permutations of the sample.
\end{enumerate}

%%%%%%%%%%%%%%%%%%%%%%%%%%%%%%%%%%%%%%%%%%%%%%%%%%%%%%%%%%%%%%%%%%%%%%%%%%%%%%%%
%%%%%%%%%%%%%%%%%%%%%%%%%%%%%%%%%%%%%%%%%%%%%%%%%%%%%%%%%%%%%%%%%%%%%%%%%%%%%%%%

\subsection{Simulation results}
~~~~The results are presented in Table \ref{simutable} with sample
sizes $n=100$, 200 and 500. The first three rows of the table are
the type I error rate for each method under different samples sizes,
which is controlled for all methods ($\mbox{Type I error rate is
between }0.03\mbox{ to }0.09$). The following rows give the power of
each method under different scenarios and sample sizes. It is clear
that as the sample size $n$ increases, the powers increase for all
the methods except the LR method under the HS, Cone, and Circle
scenarios because of the nonlinear associations of these
scenarios.\\
%%%%%%%%%%%%%%%%%%%%%%%%%%%%%%%%%%%%%%%%%%%%%%%%%%%%%%%%%%%%%%%%%%%%%%%%%%%%%%%%
\indent When the data are generated from bivariate normal
distribution, the LR method has the highest power. This is expected
because the LR method is theoretically the most powerful test under
this scenario. The ES
and DPM tests are the second best among other comparing tests.\\
%%%%%%%%%%%%%%%%%%%%%%%%%%%%%%%%%%%%%%%%%%%%%%%%%%%%%%%%%%%%%%%%%%%%%%%%%%%%%%%%
\indent The DPM test outperforms all other methods when data are
generated from the HS and the W shapes. Under the Cone shape data,
the HHG and the DPM tests both perform well. For the Circle design,
the HHG, DDP, and DPM tests all have power greater than 0.9 starting
from small sample sizes, and the ES and MIC have lower power.\\
%%%%%%%%%%%%%%%%%%%%%%%%%%%%%%%%%%%%%%%%%%%%%%%%%%%%%%%%%%%%%%%%%%%%%%%%%%%%%%%%
\indent In summary, the LR method is able to capture linear
association but loses power in the nonlinear cases. The ES method is
able to capture linear and nonlinear associations, but loses power
in some of the nonlinear cases. The HHG and DDP methods both have
high power in testing of nonlinear associations, but lose power in
the linear association, especially the HHG method. The MIC method is
a relatively conservative test compared to all other methods, and
this problem is discussed by \cite{ComMIC}. The proposed method not
only shows the ability to capture the linear association, but is
also powerful for detecting nonlinear associations in the simulation
study.
\begin{table}[h]
\caption{Power of each test (columns) for each simulation settings
and sample size $n$ (rows). A $*$ indicates that the power is
significantly different than the power of DPM
test.}\label{simutable}
\begin{center}
\scalebox{0.95}{
\begin{tabular}
{|c|c|c|c|c|c|c|c|} \hline Type & $n$ & LR & ES & HHG & DDP & MIC & DPM\\
\hline Null              & 100 & 0.06       & 0.04       & 0.03       & 0.02       & 0.03       & 0.03\\
                         & 200 & 0.06       & 0.05       & 0.03       & 0.02       & 0.07       & 0.05\\
                         & 500 & 0.09       & 0.08       & 0.05       & 0.09       & 0.02       & 0.09\\
\hline BVN               & 100 & $0.57^{*}$ & 0.49       & $0.24^{*}$ & $0.38^{*}$ & $0.13^{*}$ & 0.43\\
                         & 200 & $0.84^{*}$ & 0.78       & $0.37^{*}$ & $0.61^{*}$ & $0.25^{*}$ & 0.76\\
                         & 500 & 0.99       & 0.99       & $0.83^{*}$ & 0.99       & $0.40^{*}$ & 0.99\\
\hline HS                & 100 & $0.11^{*}$ & $0.22^{*}$ & 0.42       & 0.39       & $0.12^{*}$ & 0.44\\
                         & 200 & $0.05^{*}$ & $0.48^{*}$ & $0.53^{*}$ & $0.60^{*}$ & $0.25^{*}$ & 0.68\\
                         & 500 & $0.10^{*}$ & $0.96    $ & $0.99    $ & 1.00       & $0.42^{*}$ & 1.00\\
\hline Cone              & 100 & $0.03^{*}$ & $0.25^{*}$ & $0.54^{*}$ & $0.33    $ & $0.17^{*}$ & 0.36\\
                         & 200 & $0.08^{*}$ & $0.56^{*}$ & 0.87       & $0.73^{*}$ & $0.37^{*}$ & 0.84\\
                         & 500 & $0.09^{*}$ & 1.00       & 1.00       & 1.00       & $0.80^{*}$ & 1.00\\
\hline W                 & 100 & $0.54^{*}$ & $0.42^{*}$ & $0.55^{*}$ & $0.75^{*}$ & $0.34^{*}$ & 0.92\\
                         & 200 & $0.84^{*}$ & $0.83^{*}$ & $0.92^{*}$ & 1.00       & $0.70^{*}$ & 1.00\\
                         & 500 & 1.00       & 1.00       & 1.00       & 1.00       & $0.98    $ & 1.00\\
\hline Circle            & 100 & $0.00^{*}$ & $0.00^{*}$ & 0.96       & 0.99       & $0.20^{*}$ & 0.99\\
                         & 200 & $0.00^{*}$ & $0.22^{*}$ & 1.00       & 1.00       & $0.37^{*}$ & 1.00\\
                         & 500 & $0.00^{*}$ & 1.00       & 1.00       & 1.00       & $0.95^{*}$ & 1.00\\ \hline
\end{tabular}
}
\end{center}
\end{table}

%%%%%%%%%%%%%%%%%%%%%%%%%%%%%%%%%%%%%%%%%%%%%%%%%%%%%%%%%%%%%%%%%%%%%%%%%%%%%%%%
%%%%%%%%%%%%%%%%%%%%%%%%%%%%%%%%%%%%%%%%%%%%%%%%%%%%%%%%%%%%%%%%%%%%%%%%%%%%%%%%
%%%%%%%%%%%%%%%%%%%%%%%%%%%%%%%%%%%%%%%%%%%%%%%%%%%%%%%%%%%%%%%%%%%%%%%%%%%%%%%%
%%%%%%%%%%%%%%%%%%%%%%%%%%%%%%%%%%%%%%%%%%%%%%%%%%%%%%%%%%%%%%%%%%%%%%%%%%%%%%%%
%%%%%%%%%%%%%%%%%%%%%%%%%%%%%%%%%%%%%%%%%%%%%%%%%%%%%%%%%%%%%%%%%%%%%%%%%%%%%%%%
%%%%%%%%%%%%%%%%%%%%%%%%%%%%%%%%%%%%%%%%%%%%%%%%%%%%%%%%%%%%%%%%%%%%%%%%%%%%%%%%

\section{Real data analysis}\label{s:real}
~~~~We compare the six methods in the simulation study on the gene
expression data set from \cite{geneexpression}. Studies of
associations between genes can be found in \cite{gene1} and
\cite{gene2}. The number of observations is $n=300$ for each gene,
and we select 94 genes on chromosome 1 after removing samples with
missing values. The objective is to test
the pairwise associations within these 94 genes. A total of $\bigl(\begin{smallmatrix} 94\\
2
\end{smallmatrix} \bigr)=4371$ hypotheses tests of independence are
performed. Because of the large number of tests, we control false
discovery rate (FDR) at the 0.05 level rather than Type I error. The
Bayesian FDR (BFDR) control procedure is applied
\citep{BayesFDR,BayesFDR3,FDRcontrol,BayesFDR2} for the DPM test,
and the Benjamini--Hochberg procedure \citep{BH} is applied for
the other methods.\\
%%%%%%%%%%%%%%%%%%%%%%%%%%%%%%%%%%%%%%%%%%%%%%%%%%%%%%%%%%%%%%%%%%%%%%%%%%%%%%%%
\indent The Cohen's $\kappa$ statistic \citep{COHEN} is used to
measure agreement between tests. The $\kappa$ statistic is
\beq\kappa=\frac{P_a-P_e}{1-P_e},\nonumber\eeq where $P_a$ is the
proportion of agreements between the two methods among the $N=4371$
tests, and $P_e$ is the theoretical proportion of agreements under
independence. Larger values of $\kappa$ represents more agreement
between the tests. The number of rejections among $N=4371$ tests and
the $\kappa$ statistics of pairwise methods are presented in Table
\ref{realresult2}.
\begin{table}[h]
\caption{Numbers of rejections (of the $N=4371$ tests), and Cohen's
$\kappa$ statistics for each pair of methods.}\label{realresult2}
\begin{center}
\begin{tabular}
{|c|c|c|c|c|c|c|c|} \hline Methods & Number of rejections & LR & ES & HHG & DDP & MIC & DPM\\
\hline LR                 & 2404 & 1.000 & 0.472 & 0.301 & 0.404 & 0.082 & 0.452\\
       ES                 & 3352 &   --  & 1.000 & 0.686 & 0.830 & 0.036 & 0.779\\
       HHG                & 3442 &   --  &   --  & 1.000 & 0.751 & 0.032 & 0.720\\
       DDP                & 3350 &   --  &   --  &   --  & 1.000 & 0.036 & 0.814\\
       MIC                & 249  &   --  &   --  &   --  &   --  & 1.000 & 0.042\\
       DPM                & 3231 &   --  &   --  &   --  &   --  &   --  & 1.000\\ \hline
\end{tabular}
\end{center}
\end{table}

\begin{figure}[h]
\setlength{\abovecaptionskip}{1ex} \centering
\includegraphics[height=10cm,width=9cm]{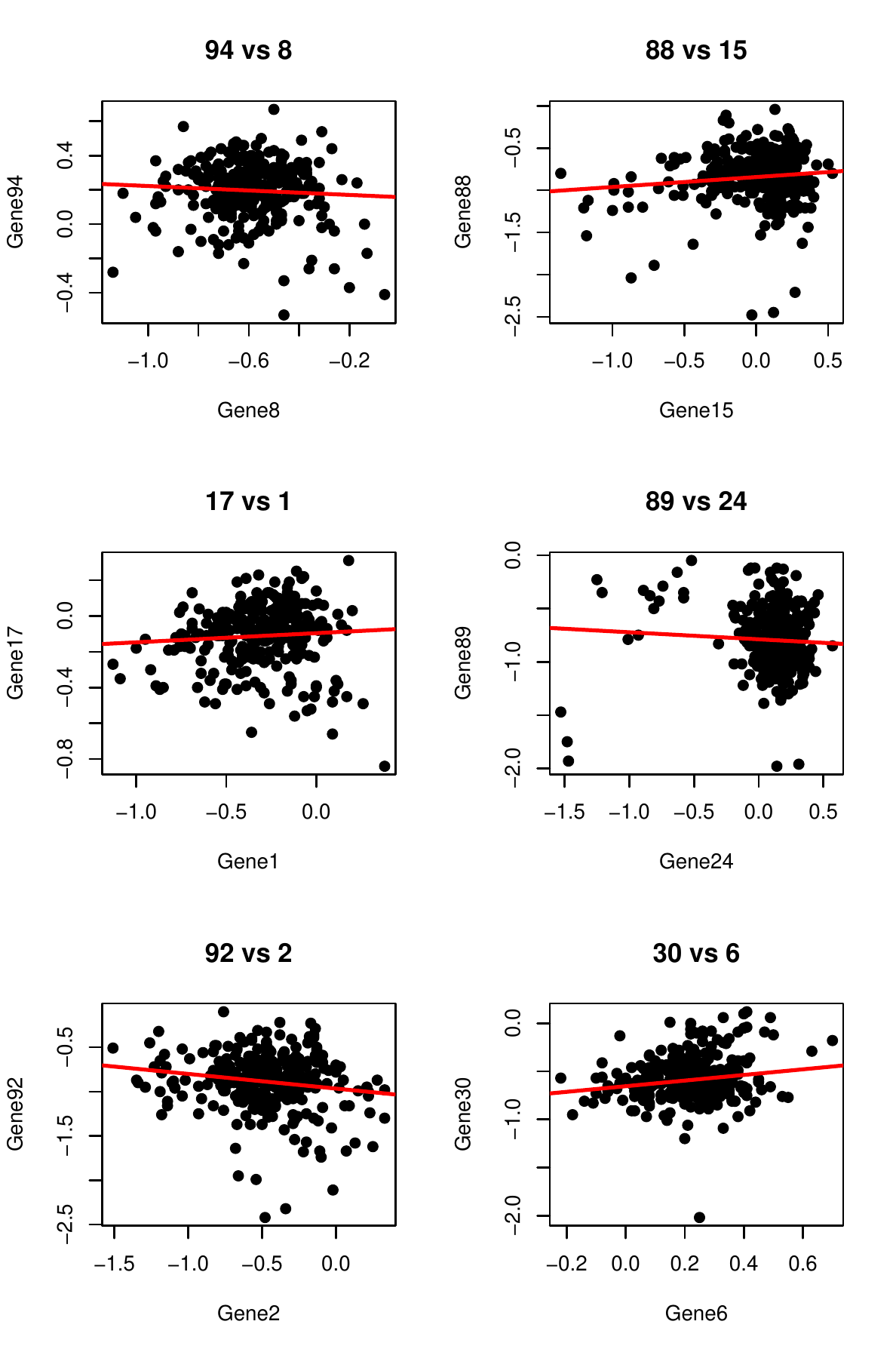}
\caption{Six pairs of genes where there are disagreements among the
tests. The red lines are the linear regression fitted
lines.}\label{4cases}
\end{figure}

\noindent The $\kappa$ statistics show that the ES, HHG, DDP, and
the DPM tests have similar testing powers in this gene expression
data sets, and the number of rejections among these tests are
similar (3231 to 3442). The LR test only captures the linear
associations between genes, and the MIC has
the lowest power as in the simulation study.\\
%%%%%%%%%%%%%%%%%%%%%%%%%%%%%%%%%%%%%%%%%%%%%%%%%%%%%%%%%%%%%%%%%%%%%%%%%%%%%%%%
\indent In Figure \ref{4cases}, we plot six pairs of genes where
there are disagreements among the tests. In the upper two plots
(gene 94 versus gene 8, and gene 88 versus gene 15), the
associations between these pairs of genes are detected by the DPM
test, but not the other tests. The figure shows that between gene 94
and gene 8, there is a horseshoe pattern of dependence, and a
nonlinear relationship between gene 88 and gene 15. In the middle
two plots (gene 17 versus gene 1, and gene 89 versus gene 24), the
ES, HHG, DDP, and DPM tests all flag associations between genes, but
not the LR and the MIC tests. The figure shows that gene 17 and gene
1 have a cone-shape association, and genes 89 and 24 have a
clustering relationship. The bottom two plots (gene 92 versus gene 2
and gene 30 versus gene 6) are the cases where only the LR, ES and
DPM tests flag associations between genes. These three tests are
powerful in testing the linear associations, and the figure shows
linear relationships between genes in these two pairs.\\

%%%%%%%%%%%%%%%%%%%%%%%%%%%%%%%%%%%%%%%%%%%%%%%%%%%%%%%%%%%%%%%%%%%%%%%%%%%%%%%%
%%%%%%%%%%%%%%%%%%%%%%%%%%%%%%%%%%%%%%%%%%%%%%%%%%%%%%%%%%%%%%%%%%%%%%%%%%%%%%%%
%%%%%%%%%%%%%%%%%%%%%%%%%%%%%%%%%%%%%%%%%%%%%%%%%%%%%%%%%%%%%%%%%%%%%%%%%%%%%%%%
%%%%%%%%%%%%%%%%%%%%%%%%%%%%%%%%%%%%%%%%%%%%%%%%%%%%%%%%%%%%%%%%%%%%%%%%%%%%%%%%
%%%%%%%%%%%%%%%%%%%%%%%%%%%%%%%%%%%%%%%%%%%%%%%%%%%%%%%%%%%%%%%%%%%%%%%%%%%%%%%%
%%%%%%%%%%%%%%%%%%%%%%%%%%%%%%%%%%%%%%%%%%%%%%%%%%%%%%%%%%%%%%%%%%%%%%%%%%%%%%%%
\section{Conclusion}\label{s:conclusion}
~~~~We propose a nonparametric Bayesian test of dependence by
calculating the Bayes factor using the Dirichlet process mixture
model and the reversible jump MCMC algorithm. We compare our method
with the linear model, distance correlation method, HHG, DDP, and
MIC in the simulation study and also in the gene expression data
sets. The simulation results show that the proposed test is
competitive in testing both linear and nonlinear relationships.\\
%%%%%%%%%%%%%%%%%%%%%%%%%%%%%%%%%%%%%%%%%%%%%%%%%%%%%%%%%%%%%%%%%%%%%%%%%%%%%%%%
\indent In the gene expression data analysis, we performed 4371
multiple testing on the gene expression data in comparing pairwise
genes. The proposed test shows similar performance with the distance
correlation, DDP, and HHG methods, and detects some cases that other
methods do not detect. It also shows that the proposed method is
powerful on both linear and nonlinear relationships in the pairwise
gene comparisons.

%%%%%%%%%%%%%%%%%%%%%%%%%%%%%%%%%%%%%%%%%%%%%%%%%%%%%%%%%%%%%%%%%%%%%%%%%%%%%%%%
%%%%%%%%%%%%%%%%%%%%%%%%%%%%%%%%%%%%%%%%%%%%%%%%%%%%%%%%%%%%%%%%%%%%%%%%%%%%%%%%
%%%%%%%%%%%%%%%%%%%%%%%%%%%%%%%%%%%%%%%%%%%%%%%%%%%%%%%%%%%%%%%%%%%%%%%%%%%%%%%%
%%%%%%%%%%%%%%%%%%%%%%%%%%%%%%%%%%%%%%%%%%%%%%%%%%%%%%%%%%%%%%%%%%%%%%%%%%%%%%%%
%%%%%%%%%%%%%%%%%%%%%%%%%%%%%%%%%%%%%%%%%%%%%%%%%%%%%%%%%%%%%%%%%%%%%%%%%%%%%%%%
%%%%%%%%%%%%%%%%%%%%%%%%%%%%%%%%%%%%%%%%%%%%%%%%%%%%%%%%%%%%%%%%%%%%%%%%%%%%%%%%

\newpage
\appendix

\section{Full conditional distributions}\label{fullcond}

%%%%%%%%%%%%%%%%%%%%%%%%%%%%%%%%%%%%%%%%%%%%%%%%%%%%%%%%%%%%%%%%%%%%%%%%%%%%%%%%
%%%%%%%%%%%%%%%%%%%%%%%%%%%%%%%%%%%%%%%%%%%%%%%%%%%%%%%%%%%%%%%%%%%%%%%%%%%%%%%%

\subsection{Full conditionals for the independent DPM
prior}\label{MCMCindep} ~~~~Let
$\mat{X}_j=\{\mat{X}_{ij}:i=1,...,N\}$, where $\mat{X}_{ij}$ is the
$i^{th}$ observation of $\mat{X}_j$ and $N$ is the number of
observations. The prior of $\mat{S}_j$ is
$\mat{S}_j\sim\mbox{IW}_{D_j}(\rho_j,W_j)$. The full conditional
distribution for each parameters under the independent DPM prior of
$\mat{X}_j$ are
\beqn\mat{\mu}_{lj}~|~\mbox{rest}&\sim&\mbox{MVN}_{D_j}[(n_{lj}\mat{\Sigma}_j^{-1}+\mat{\Omega}_j^{-1})^{-1}\mat{\Sigma}_j^{-1}(\sum_{i:g_{ij}=l}\mat{X}_{ij}),(n_{lj}\mat{\Sigma}_j^{-1}+\mat{\Omega}_j^{-1})^{-1}]\nonumber\\
     \mat{S}_j~|~\mbox{rest}&\sim&\mbox{IW}[N+K+\rho_j,A]\nonumber\\
         P(r=r_m~|~\mbox{rest})&=&\frac{\prod_{i=1}^N\phi_j(\mat{X}_{ij}~|~\mat{\mu}_{g_{ij}j},r_m\mat{S}_j)\prod_{l=1}^K\phi_j(\mat{\mu}_{lj}~|~\mat{0},(1-r_m)\mat{S}_j)\frac{1}{n_r}}{\sum_{q=1}^{n_r}\left[\prod_{i=1}^N\phi_j(\mat{X}_{ij}~|~\mat{\mu}_{g_{ij}j},r_q\mat{S}_j)\prod_{l=1}^K\phi_j(\mat{\mu}_{lj}~|~\mat{0},(1-r_q)\mat{S}_j)\frac{1}{n_r}\right]}\nonumber\\
         P(g_{ij}=l~|~\mbox{rest})&=&\frac{\phi_j(\mat{X}_{ij}~|~\mat{\mu}_{lj},\mat{\Sigma}_j)w_{lj}}{\sum_{s=1}^{K}\left[\phi_j(\mat{X}_{ij}~|~\mat{\mu}_{sj},\mat{\Sigma}_j)w_{sj}\right]}\nonumber\\
         v_{lj}~|~\mbox{rest}&\sim&\mbox{Beta}\left[\sum_{i=1}^NI(g_{ij}=l)+1,\sum_{i=1}^NI(g_{ij}>l)+d_j\right]\nonumber\\
         d_j~|~\mbox{rest}&\sim&\mbox{Gamma}\left[K+a-1,b-\sum_{l=1}^{K-1}\mbox{log}(1-v_{lj})\right],\nonumber\eeqn
where $l=1,...,K$, $m=1,...,n_r$, $i=1,...,N$,
$A=r^{-1}\sum_{i=1}^N(\mat{X}_{ij}-\mat{\mu}_{g_{ij}j})(\mat{X}_{ij}-\mat{\mu}_{g_{ij}j})^{T}+(1-r)^{-1}\sum_{l=1}^{K}\mat{\mu}_{lj}\mat{\mu}_{lj}^{T}+\rho_jW_j$,
$n_{lj}=\sum_{i=1}^{N}I(g_{ij}=l)$, $g_{ij}$ is the cluster label of
the $i^{th}$ observation, $n_r$ is the number of discrete $r$
values, $\phi_j$ is the $D_j$-dimensional multivariate normal
density function, and $(a,b)$ is the tunning parameter of the
stick-breaking algorithm.

%%%%%%%%%%%%%%%%%%%%%%%%%%%%%%%%%%%%%%%%%%%%%%%%%%%%%%%%%%%%%%%%%%%%%%%%%%%%%%%%
%%%%%%%%%%%%%%%%%%%%%%%%%%%%%%%%%%%%%%%%%%%%%%%%%%%%%%%%%%%%%%%%%%%%%%%%%%%%%%%%

\subsection{Full conditionals for the joint DPM
prior}\label{MCMCjoint} ~~~~Let $\mat{X}=\{\mat{X}_{i}:i=1,...,N\}$,
where $\mat{X}_{i}$ is the $i^{th}$ observation of $\mat{X}$ and $N$
is the number of observations. The prior of $S$ is
$S\sim\mbox{IW}_{D_1+D_2}(\rho,W)$. The full conditional
distribution for each parameters under the joint DPM prior of
$\mat{X}$ are
\beqn\mat{\mu}_{l}~|~\mbox{rest}&\sim&\mbox{MVN}_{D_1+D_2}[(n_l\mat{\Sigma}^{-1}+\mat{\Omega}^{-1})^{-1}\mat{\Sigma}^{-1}(\sum_{i:g_{i}=l}\mat{X}_{i}),(n_l\mat{\Sigma}^{-1}+\mat{\Omega}^{-1})^{-1}]\nonumber\\
      \mat{S}~|~\mbox{rest}&\sim&\mbox{IW}[N+K+\rho,B]\nonumber\\
         P(r=r_m~|~\mbox{rest})&=&\frac{\prod_{i=1}^N\phi(\mat{X}_{i}~|~\mat{\mu}_{g_{i}},r_m\mat{S})\prod_{l=1}^K\phi(\mat{\mu}_{l}~|~\mat{0},(1-r_m)\mat{S})\frac{1}{n_r}}{\sum_{q=1}^{n_r}\left[\prod_{i=1}^N\phi(\mat{X}_{i}~|~\mat{\mu}_{g_{i}},r_q\mat{S})\prod_{l=1}^K\phi(\mat{\mu}_{l}~|~\mat{0},(1-r_q)\mat{S})\frac{1}{n_r}\right]}\nonumber\\
         P(g_{ij}=l~|~\mbox{rest})&=&\frac{\phi_j(\mat{X}_{ij}~|~\mat{\mu}_{lj},\mat{\Sigma}_j)w_{lj}}{\sum_{s=1}^{K}\left[\phi_j(\mat{X}_{ij}~|~\mat{\mu}_{sj},\mat{\Sigma}_j)w_{sj}\right]}\nonumber\\
         v_{l}~|~\mbox{rest}&\sim&\mbox{Beta}\left[\sum_{i=1}^NI(g_{i}=l)+1,\sum_{i=1}^NI(g_{i}>l)+d\right]\nonumber\\
         d~|~\mbox{rest}&\sim&\mbox{Gamma}\left[K+a-1,b-\sum_{l=1}^{K-1}\mbox{log}(1-v_{l})\right],\nonumber\eeqn
where $l=1,...,K$, $m=1,...,n_r$, $i=1,...,N$,
$B=r^{-1}\sum_{i=1}^N(\mat{X}_{i}-\mat{\mu}_{g_{i}})(\mat{X}_{i}-\mat{\mu}_{g_{i}})^{T}+(1-r)^{-1}\sum_{l=1}^{K}\mat{\mu}_{l}\mat{\mu}_{l}^{T}+\rho
W$, $n_{l}=\sum_{i=1}^{N}I(g_{i}=l)$, $g_{i}$ is the cluster label
of the $i^{th}$ observation, $n_r$ is the number of discrete $r$
values, $\phi$ is the $D$-dimensional multivariate normal density
function, and $(a,b)$ is the tunning parameter of the stick-breaking
algorithm.

%%%%%%%%%%%%%%%%%%%%%%%%%%%%%%%%%%%%%%%%%%%%%%%%%%%%%%%%%%%%%%%%%%%%%%%%%%%%%%%%
%%%%%%%%%%%%%%%%%%%%%%%%%%%%%%%%%%%%%%%%%%%%%%%%%%%%%%%%%%%%%%%%%%%%%%%%%%%%%%%%
%%%%%%%%%%%%%%%%%%%%%%%%%%%%%%%%%%%%%%%%%%%%%%%%%%%%%%%%%%%%%%%%%%%%%%%%%%%%%%%%
%%%%%%%%%%%%%%%%%%%%%%%%%%%%%%%%%%%%%%%%%%%%%%%%%%%%%%%%%%%%%%%%%%%%%%%%%%%%%%%%
%%%%%%%%%%%%%%%%%%%%%%%%%%%%%%%%%%%%%%%%%%%%%%%%%%%%%%%%%%%%%%%%%%%%%%%%%%%%%%%%
%%%%%%%%%%%%%%%%%%%%%%%%%%%%%%%%%%%%%%%%%%%%%%%%%%%%%%%%%%%%%%%%%%%%%%%%%%%%%%%%

\section{Test of independence by E-statistics}\label{APP2}
~~~~The test of independence by E-statistics, which calculates the
distance covariance measures (dCov), was first introduced by
\cite{dCov}. The dCov between two random variables (or vectors)
$\mat{X}_1\in\mathbb{R}^p$ and $\mat{X}_2\in\mathbb{R}^q$ with
finite first moments is the nonnegative number defined as
\beq\label{dcovformula}\mathcal{V}^2(\mat{X}_1,\mat{X}_2)=\|f(\mat{X})-f_1(\mat{X}_1)f_2(\mat{X}_2)\|^2_{w},\eeq
where $\|\cdot\|^2_w$ is the $L_2$-norm with weight function $w$.
The $w$ is described more details in \cite{dCov}, and in this
article, we use the identical $w$ as suggested. The empirical
distance covariance of $n$ observed samples
$\mathcal{V}_n^2(\mat{X}_1,\mat{X}_2)$ is also defined in
\cite{dCov}. A test statistic $T(\mat{X}_1,\mat{X}_2,p,n)$ that
rejects the null hypothesis that two random variables (or vectors)
if
$$\frac{n\mathcal{V}_n^2(\mat{X}_1,\mat{X}_2)}{S_2}>(\Phi^{-1}(1-\alpha/2))^2$$
has an asymptotic significance level at most $\alpha$, and
$$S_2=\frac{1}{n^2}\sum_{k,l=1}^{n}|\mat{X}_{1k}-\mat{X}_{1l}|_p\frac{1}{n^2}\sum_{k,l=1}^{n}|\mat{X}_{2k}-\mat{X}_{2l}|_q,$$
where $|\cdot|_r$ is the $L_r$-norm, and $\Phi$ is the standard
normal distribution. However, the test decision based on $\Phi$ is
quite conservative for many distributions, so the testing decision
in this article is determined by 300 permutation samples under the
null hypothesis with Type I error rate $p=0.05$ level under each
data set. The R package "$energy$" with function
"$indep.test$" is used in the analysis.\\
%%%%%%%%%%%%%%%%%%%%%%%%%%%%%%%%%%%%%%%%%%%%%%%%%%%%%%%%%%%%%%%%%%%%%%%%%%%%%%%%
\indent A distribution-free version of distance covariance was also
introduced in \citep{rankdCov}, which uses the ranks of the
observations instead of the values. In this article, the
distribution-free version of dCov performs similar to the original
version, so we only present the original version of the dCov.

%%%%%%%%%%%%%%%%%%%%%%%%%%%%%%%%%%%%%%%%%%%%%%%%%%%%%%%%%%%%%%%%%%%%%%%%%%%%%%%%
%%%%%%%%%%%%%%%%%%%%%%%%%%%%%%%%%%%%%%%%%%%%%%%%%%%%%%%%%%%%%%%%%%%%%%%%%%%%%%%%
%%%%%%%%%%%%%%%%%%%%%%%%%%%%%%%%%%%%%%%%%%%%%%%%%%%%%%%%%%%%%%%%%%%%%%%%%%%%%%%%
%%%%%%%%%%%%%%%%%%%%%%%%%%%%%%%%%%%%%%%%%%%%%%%%%%%%%%%%%%%%%%%%%%%%%%%%%%%%%%%%
%%%%%%%%%%%%%%%%%%%%%%%%%%%%%%%%%%%%%%%%%%%%%%%%%%%%%%%%%%%%%%%%%%%%%%%%%%%%%%%%
%%%%%%%%%%%%%%%%%%%%%%%%%%%%%%%%%%%%%%%%%%%%%%%%%%%%%%%%%%%%%%%%%%%%%%%%%%%%%%%%

\section{Heller-Heller-Gorfine test of association based on Euclidean
distance metric}\label{APP3}~~~~This test was first introduced by
\cite{HHG}. The test is based on the pairwise distances within
$\mat{X}_1$ and $\mat{X}_2$ respectively. Let the pairwise distances
within $\mat{X}_j$, $j=1,2$, denoted as
$\{d(\mat{X}_{ij},\mat{X}_{i^{'}j}):i,i^{'}\in\{1,...,n\}\}$, where
$\mat{X}_{ij}$ is the $i^{th}$ observation in $\mat{X}_j$, and
$d(\cdot,\cdot)$ is assigned to be the Euclidean distance metric in
this article. The idea is to first randomly select two samples $i$
and $i^{'}$ in each of $\mat{X}_1$ and $\mat{X}_2$, and then use the
distances $d(\mat{X}_{i1},\mat{X}_{i^{'}1})$ and
$d(\mat{X}_{i2},\mat{X}_{i^{'}2})$ as the references to construct a
$2\times 2$ contingency table among the remaining $n-2$ samples.
Then the likelihood ratio test of independence for summarizing this
table denoted as $S(i,i^{'})$ gives test statistic
$$T=\sum_{i=1}^{n}\sum_{i^{'}=1,i^{'}\neq i}^{n}S(i,i^{'}).$$ The
0.05 Type I error rate is controlled by 300 permutation samples
under the null hypothesis of each data set. The R package "$HHG$"
with function "$hhg.test$" is used in the
analysis.\\
%%%%%%%%%%%%%%%%%%%%%%%%%%%%%%%%%%%%%%%%%%%%%%%%%%%%%%%%%%%%%%%%%%%%%%%%%%%%%%%%
\indent A distribution-free version of HHG test was suggested in
\citep{HHG} for comparison. We found that in this article the
results are similar to the original version of HHG test. Therefore,
we only present the original version of the HHG results.

%%%%%%%%%%%%%%%%%%%%%%%%%%%%%%%%%%%%%%%%%%%%%%%%%%%%%%%%%%%%%%%%%%%%%%%%%%%%%%%%
%%%%%%%%%%%%%%%%%%%%%%%%%%%%%%%%%%%%%%%%%%%%%%%%%%%%%%%%%%%%%%%%%%%%%%%%%%%%%%%%
%%%%%%%%%%%%%%%%%%%%%%%%%%%%%%%%%%%%%%%%%%%%%%%%%%%%%%%%%%%%%%%%%%%%%%%%%%%%%%%%
%%%%%%%%%%%%%%%%%%%%%%%%%%%%%%%%%%%%%%%%%%%%%%%%%%%%%%%%%%%%%%%%%%%%%%%%%%%%%%%%
%%%%%%%%%%%%%%%%%%%%%%%%%%%%%%%%%%%%%%%%%%%%%%%%%%%%%%%%%%%%%%%%%%%%%%%%%%%%%%%%
%%%%%%%%%%%%%%%%%%%%%%%%%%%%%%%%%%%%%%%%%%%%%%%%%%%%%%%%%%%%%%%%%%%%%%%%%%%%%%%%

\section{Distribution-free tests of association based on data derived partitions}\label{APP4}
~~~~This test was first introduced in \citep{DDP}, which is designed
for testing two univariate random variables (i.e. $D_1=D_2=1$). The
idea follows the HHG test but with different ways of forming the
contingency tables. The data values are now used directly instead of
using the distances. In forming a $2\times 2$ contingency table, one
sample point is randomly selected as the reference, and then a
$2\times 2$ contingency table can be constructed and a test
statistic of this table is calculated. The same procedure can be
applied to form $m\times m$ contingency tables ($m>2$) with randomly
selected $m-1$ data values as references. More specifically, the
$m\times m$ contingency table is defined by the range
$(-\infty,X^{*}_{1(1)})$, $(X^{*}_{1(2)},X^{*}_{1(3)})$,...,
$(X^{*}_{1(m-1)},\infty)$ in $\mat{X}_1$, and
$(-\infty,X^{*}_{2(1)})$, $(X^{*}_{2(2)},X^{*}_{2(3)})$,...,
$(X^{*}_{2(m-1)},\infty)$ in $\mat{X}_2$, where $X^{*}_{j(r)}$ is
the $r^{th}$ ordered selected observation in $\mat{X}_j$, $j=1,2$.
In this article, the summation of the likelihood ratio test
statistics with each $3\times 3$ ($m=3$) contingency table is used
as the test statistics. This setting was shown to perform the best
in most of the scenarios in \citep{DDP}. The testing decision is
again based on 300 permutation samples under the null hypothesis for
each data sets controlled under 0.05 Type I error rate in this
study. The R package "$HHG$" with function "$xdp.test$" is used in
this article.

%%%%%%%%%%%%%%%%%%%%%%%%%%%%%%%%%%%%%%%%%%%%%%%%%%%%%%%%%%%%%%%%%%%%%%%%%%%%%%%%
%%%%%%%%%%%%%%%%%%%%%%%%%%%%%%%%%%%%%%%%%%%%%%%%%%%%%%%%%%%%%%%%%%%%%%%%%%%%%%%%
%%%%%%%%%%%%%%%%%%%%%%%%%%%%%%%%%%%%%%%%%%%%%%%%%%%%%%%%%%%%%%%%%%%%%%%%%%%%%%%%
%%%%%%%%%%%%%%%%%%%%%%%%%%%%%%%%%%%%%%%%%%%%%%%%%%%%%%%%%%%%%%%%%%%%%%%%%%%%%%%%
%%%%%%%%%%%%%%%%%%%%%%%%%%%%%%%%%%%%%%%%%%%%%%%%%%%%%%%%%%%%%%%%%%%%%%%%%%%%%%%%
%%%%%%%%%%%%%%%%%%%%%%%%%%%%%%%%%%%%%%%%%%%%%%%%%%%%%%%%%%%%%%%%%%%%%%%%%%%%%%%%

\section{Maximal information coefficient for measuring dependence of two
variables}\label{APP5}~~~~The Maximal Information Coefficient (MIC)
method is first introduced by \cite{MIC}. The intuition is that if a
relationship exists between two univariate random variables, then a
grid (a square) can be drawn on the scatter-plot of these two
variables which can partition the data to capture the relationship.
The method explores all size of grids up to a maximal grid
resolution. For grid size $x$-by-$y$, the largest achievable
normalized mutual information (MI) is denoted as $m_{xy}$
$$m_{xy}=\mbox{max}\{I_{xy}\}/\mbox{log}(\mbox{min}\{x,y\}),$$ where
computation of $I_{xy}$ can be found in \citep{MI}, and the MIC is
the maximum of $m_{xy}$ over all pair $(x,y)$ such that $xy<B$,
where $B$ depends on the sample size $n$. In this article, we use
$B=n^{0.6}$ as suggested in \cite{MIC}, and the p-value is
calculated from the p-value table given in
$www.exploredata.net/Downloads/P-Value-Tables$. The R package
"$minerva$" with function "$mine$" is used in this article.

% Start a new page
%\newpage
%\appendix
%\section{}
% Start a new page

\newpage
{\singlespacing
\bibliography{bib}
\bibliographystyle{agsm}
}

% End of the article
\end{document}